\documentclass[]{aa}  
\usepackage{graphicx}
\usepackage{txfonts}
\usepackage[]{hyperref}
\usepackage{todonotes}

\newcommand{\mct}{SB\,744}
\newcommand{\xtgrid}{{\sc XTgrid}}
\renewcommand{\sun}{$_{\odot}$}
\newcommand{\pbfour}{\ion{Pb}{iv}}
\newcommand{\pbthree}{\ion{Pb}{iii}}
\begin{document} 

\title{The first heavy-metal hot subdwarf composite binary \mct}
\author{P. N\'emeth \inst{1,2} \and
          J. Vos \inst{1,3} \and
          F. Molina \inst{3,4} \and
          A. Bastian \inst{3}
          }
\institute{Astronomical Institute of the Czech Academy of Sciences, CZ-251\,65, Ond\v{r}ejov, Czech Republic\\
\email{peter.nemeth@astroserver.org} \and
Astroserver.org, F\H{o} t\'{e}r 1, 8533 Malomsok, Hungary \and
Institut f\"{u}r Physik und Astronomie, Universit\"{a}t Potsdam, Karl-Liebknecht-Str. 24/25, 14476, Golm, Germany \and
Laboratorio LCH, Centro de Investigación en Química Sostenible (CIQSO), Universidad de Huelva, Huelva 21007, Spain 
}
\date{Received -; accepted -}

\abstract
{}
{A radial velocity follow-up of the long-period sdOB+G1V type spectroscopic binary \mct\ revealed strong lines of fluorine and lead in the optical spectrum of the sdOB star and subsolar metallicity in the G1V companion. 
With high quality observations and Gaia astrometric data we aim at measuring the chemical composition and Galactic kinematics of the system to put it in context with known populations of hot subdwarfs. 
Such binary systems have high potential, as they give insights into the late stages of binary evolution as well as into the mysterious formation of stripped core helium burning stars.}
{We have analyzed the optical spectra with homogeneous atmospheric models to derive surface parameters of the binary members from a direct wavelength space decomposition and independently measured the atmospheric properties of the cool companion. 
The two independent methods reached consistent results, which, amended with constraints from spectral energy distributions provided a subdwarf mass. 
The Gaia astrometry allowed us to derive the Galactic kinematics of the system.}
{\mct\ turned out to be an old, Population II system, that has gone through dramatic events. 
The hot subdwarf star belongs to the heavy-metal subclass of sdOB stars and we report super-solar abundances of lead, based on Pb\,{\sc iii/iv} lines. 
The He abundance of the hot subdwarf is the lowest among the known heavy-metal sdOB stars.
The presence of fluorine implies that \mct\ was once a hierarchical triple system and the inner binary has merged in the near past. 
As an alternative scenario, single-star evolution through late core helium flash and atmospheric mixing can also produce the observed fluorine abundances.
The atmospheric metal over-abundances currently observed are perhaps the results of a combination of mixing processes during formation and radiative support.}
{}
\keywords{Stars: abundances, atmospheres, individual: SB 744, evolution, chemically peculiar; subdwarfs; binaries: spectroscopic}

\maketitle
%

\section{Introduction}
Hot subdwarf stars are 0.5 M$_\odot$ core helium burning stars on the extreme horizontal branch (EHB).
Unlike normal horizontal branch stars they retain only a very thin hydrogen envelope after their evolution on the red giant branch (RGB). 
The envelope is lost under unclear circumstances, precisely by the time they ignite helium burning in the core. 
The precise timing and high binary fraction hints a yet unknown connection between mass loss and binary evolution (\citealt{heber09, heber16}). 

Binary population synthesis models outlined three major channels to form hot subdwarf stars: The Roche-lobe overflow (RLOF), the common-envelope (CE) and the binary white dwarf merger evolution, that are all related to binarity \citep{Han2002}. 
However, of the three channels, only the white dwarf merger channel predicts single hot subdwarfs \citep{zhang12a}. 
As an alternative to binary evolution, single star evolution models with a delayed core He-flash can also produce hot subdwarfs and reproduce their observed properties \citep{Marcelo08}. 

The spectral properties of hot subdwarfs show a nearly continuous sequence from 25\,000\,K B-type subdwarfs (sdB) to the hottest sdO type stars exceeding 40\,000\,K. 
However, there are parallel sequences well separated by helium abundance, and, along these sequences, multiple groups can be identified (\citealt{edelmann03, nemeth12}). 
This granulation in parameter space correlates with other properties, such as binarity and most likely indicate different formation pathways that lead to different sub-classes of hot subdwarfs. 
The spectra of the intermediate helium-rich sub-class show a mix of He\,{\sc i} and {\sc ii} lines (sdOB type) and their location in the parameter space differ from the canonical picture of sdB and sdO stars. 
A relation between sdB and sdO stars can be outlined by stellar evolution, where sdB stars are the progeny of post-EHB sdO stars. 
The similar binary properties of sdB and sdO types confirm this evolutionary link.
However, the sdOB subclass shows a very low binary fraction, implying a different formation history. 
This hypothesis got further support when following \citet{Naslim11} a set a studies found heavy-metal over-abundances in several sdOB stars (\citealt{Jeffery12, Naslim13, Jeffery13, Jeffery17}). 
The trans-iron metal abundances reach over 10\,000 times the solar values in these stars. 
Such heavy metals are produced by the $s$-process during intermediate evolution within the He-shell burning environments of low metallicity stars. 
However, whether these sdOB stars have extra amounts of heavy-metals or the observed abundances are the result of diffusion, which places metals selectively into a thin photospheric layer, is not yet clear. 

Among the 11 heavy-metal subdwarfs currently cataloged \citep{Naslim20}, two pulsators, LSIV-14$^\circ$116 and Feige\,48 have been identified \citep{Dorsch20}. 
The two stars show strikingly similar temperature, gravity and pulsation properties. 
Interestingly, two patterns seem to emerge: cooler and pulsating stars show Y, Zr, Sr overabundances, while hotter and non-pulsating heavy-metal subdwarfs show only Pb enrichment (\citealt{menchero20, Naslim20}).
The two different heavy-metal abundance patters are informally referred to as the Sr and Pb groups. 
The $s$-process produces Sr peak elements in high metallicity stars and neutron capture can go all the way to Pb in low metallicity stars \citep{travaglio04}. 

Detailed quantitative spectral analyses of double-lined composite spectrum binaries have been overlooked in the past decades. 
In such systems the precision is limited because one must fit two stars in a single spectrum and the orbital period is too long to apply Fourier disentangling methods (e.g. \citealt{hadrava09}). 
Accuracy also suffers from large systematic shifts due to parameter correlations among the binary members. 
At the same time, subdwarf binary systems with main sequence companions are perfectly suitable for a direct, wavelength-space decomposition as pioneered by \citet{sturm94} thanks to the very different components. 
Using this method, \citet{nemeth12} have resolved 29 composite systems. 
Recent updates of the method enabled to perform a more consistent disentangling and recover the radial velocities of the members \citep{Reed20}. 

Atomic diffusion is a complex process and a key to understand stellar evolution on the horizontal branch and the horizontal branch morphology of globular clusters. 
The observed surface abundances depend on the conditions at formation (initial mass, abundances after the giant branch) and the diffusive equilibrium abundances along the subsequent evolution, hence the age of the system. 
This requires a simultaneous optimization of multiple (time-dependent) model parameters, which is a complicated task for single stars. 
The binary nature of \mct\ and the presence of Pb and F in its atmosphere provide additional constraints, and make the system very important to move the field forward.

\mct\ (also MCT\,0146-2651) is a bright (V=12.31 mag) hot subdwarf composite spectrum binary. 
It was first identified by \cite{Setteback71} as the 744th object in their catalog of early type stars near the south Galactic Pole. 
It is a relatively well studied object, its first quantitative spectral analyses date back to the 1980s (\citealp{Heber86, Unglaub89}). 
\cite{Unglaub90} performed a detailed analysis by removing the contributions of the cool companion and derived $T_{\rm eff}=36\,000\pm3000$\,K, $\log{g}=5.7\pm0.3$\,cm\,s$^{-2}$ and $\log(n{\rm He}/n{\rm H})=-1.5\pm0.2$ for the sdOB primary. 
The physical binary nature of the system could not be confirmed by radial velocity measurements back then, but the low star density in the halo and the similarity to other composite hot subdwarf binaries made it very likely to be a physical pair. 
\cite{Unglaub90} not only derived the atmospheric parameters correctly, but their estimated distance of 499 pc to the system is very close to the most recent measurements. 
The system has appeared in various surveys over the next two decades and was classified as an sdO7:He1+F/G type composite spectrum hot subdwarf (\citealp{Lamontagne00, Stark06}), but, in general, dedicated studies avoided \mct\ for nearly 30 years. 
\cite{Vos2018a} started a systematic long-term radial velocity follow program for selected hot subdwarf composite spectrum binaries, among them with \mct,
to investigate their orbital period -- mass ratio correlation. 
\cite{Vos2019a} found that the system has a $P=768\pm11$\,d orbit and refined the classification of the secondary component to G1V. 
To improve the precision of radial velocity work we fitted the observations and computed synthetic spectra to be used as templates in the cross correlation. 
These templates revealed lines of lead and fluorine in the residuals of the composite fits \citep{JVos19}. 

This article reports that the sdOB star in \mct\ is a heavy-metal hot subdwarf star.
The mere existence of this system raises questions: 
\begin{itemize}
    \item If the F and Pb overabundances are caused by binary evolution, does it work the same way in different evolution channels (e.g.: RLOF, CE evolution)? 
    \item However, if the observed abundances are the result of pure atmospheric effects, why not all such sdOB stars with very similar atmospheric properties show similar overabundances?
\end{itemize}

\section{Spectroscopy}

\subsection{Observations}
\mct\ is part of a long-term radial-velocity monitoring observing program on the Very Large Telescope (VLT) at the European Southern Observatory (ESO Program ID: 088.D-0364, 093.D-0629 and 0104.D-0135). 
It has been observed with the Ultraviolet-Visible Echelle Spectrograph (UVES) 14 times between 2013 and 2020. 
The spectra cover the optical spectral range between 3750\,\AA\ and 9000\,\AA\ at a mean signal-to-noise ratio (SNR) of 50.
The observations were performed in service mode and data reduction was done using standard procedures in {\sl EsoReflex} \citep{freudling13}.

Ultraviolet spectra are also available in the MAST archive. 
Low-resolution spectra have been obtained by the International Ultraviolet Explorer (IUE) satellite with both the short and long wavelength apertures covering the ultraviolet between $1150$ and $3200$\,\AA. 

\subsection{Analysis of the composite spectrum with \xtgrid}
The available observations of \mct\ do not sample the entire orbital period sufficiently, therefore a wavelength space decomposition is more suitable to investigate the components. 
Then, spectral decomposition of double-lined binaries require one to fit the linear combination of two synthetic spectra to observations. 
This approach is able to overcome heavy line blending and faster to implement. 
Such a procedure is available in the \xtgrid\ \citep{nemeth12} spectral analysis code.
We updated the code to work with high-resolution data, disentangle radial velocities and maintain a better consistency of the wavelength dependent flux contribution (dilution factor) of the components to the composite spectrum.
We also amended the spectral analysis with a SED fitting procedure. 
In this approach the surface parameters of the components are derived from the spectra, and the SED constraints are used only to improve the precision of the dilution factor.

The procedure applies {\sc Tlusty/Synspec} non-Local Thermodynamic Equilibrium (non-LTE) model atmospheres (\citealt{hubeny17, lanz07}) and corresponding synthetic spectra for the hot subdwarf. 
The spectral models include the first 30 elements of the periodic table and in addition: Ga, Ge, Sr, Y, Zr, Tc, Xe and Pb. 
Important elements for the atmospheric structure, such as H and He, C, N, O, Ne, Si, P, S, Fe and Ni are included in non-LTE, while the remaining elements are calculated assuming LTE conditions in the spectral synthesis. 
Synthetic spectra for the cool companion are extracted from the BOSZ spectral library \citep{bohlin17}, that were calculated with the {\sc Atlas9} \citep{Kurucz1979} code in LTE.
To improve the internal consistency of our analysis we updated the decomposition procedure in \xtgrid\ to force a more consistent flux contribution across the entire optical range.
Recent updates in \xtgrid\ allow to add new elements to the hot subdwarf model during runtime and fit spectra from different instruments together.

We started the fits using the atmospheric and binary parameters of \mct\ derived by \cite{Vos2018a}. 
The spectral properties, the flux contributions and radial velocities of both stars were optimized simultaneously and iteratively to match the observation. 
\xtgrid\ fits the effective temperature, surface gravity and projected rotation velocity of both components. 
Abundances of the hot subdwarf were determined individually for each element and the scaled solar metallicity (where differences in [M/H] are analogous to differences in [Fe/H]) with alpha element enhancement corrections were fitted for the cool companion. 
Microturbulent velocity is neglected for the subdwarf and included as a constant value (2\,km\,s$^{-1}$) in the synthetic spectra of the BOSZ library. 
We neglected macroturbulence in both components.
\xtgrid\ optimizes 38 parameters for the hot subdwarf, 6 parameters for the cool companion and, in addition, the dilution factor for the flux contributions. 
The dilution is updated iteratively along with the other free parameters following the steepest gradient of the global $\chi^2$ until the fit converges on the best solution.

Although the radial velocities of the components can be determined from the individual observations directly, the luminosities are not easy to estimate without knowing the precise radii and surface gravities. 
This can be improved when precise distances to the binaries are used, such as distances from the Gaia Data Release 2 or 3.
However, in spite of the known distance to the system, $d=483^{+16}_{-15}$ pc \citep{Bailer18}, strong correlations between metallicity, flux contribution and projected rotation are observed in the current data set. 
These correlations can be easily resolved in eclipsing binaries, where relative parameters can be measured, here we could rely only on the relatively high SNR of the observations and the wide coverage across the optical spectral range, that includes several major spectral features from both components. 
The synthetic spectral templates resulting from the simultaneous modeling of the components were also used to increase the precision of radial velocity cross-correlation measurements of the hot subdwarf. 

Parameter errors were calculated in two dimensions for $T_{\rm eff}$ and $\log{g}$ to include their correlations. 
For abundances and binary parameters the error calculations are done in one dimension, independently to the upper and lower error bars. 
Then the asymmetric errors are recalculated to symmetric ones. 
The low relative contributions of weak metal lines from the hot subdwarf to the composite spectrum make parameter determination more difficult from composite spectra, and is reflected in the error bars. 

\begin{figure*}[!ht]
\centering
  \begin{minipage}{\textwidth}
  \centering
  \includegraphics[width=0.95\linewidth]{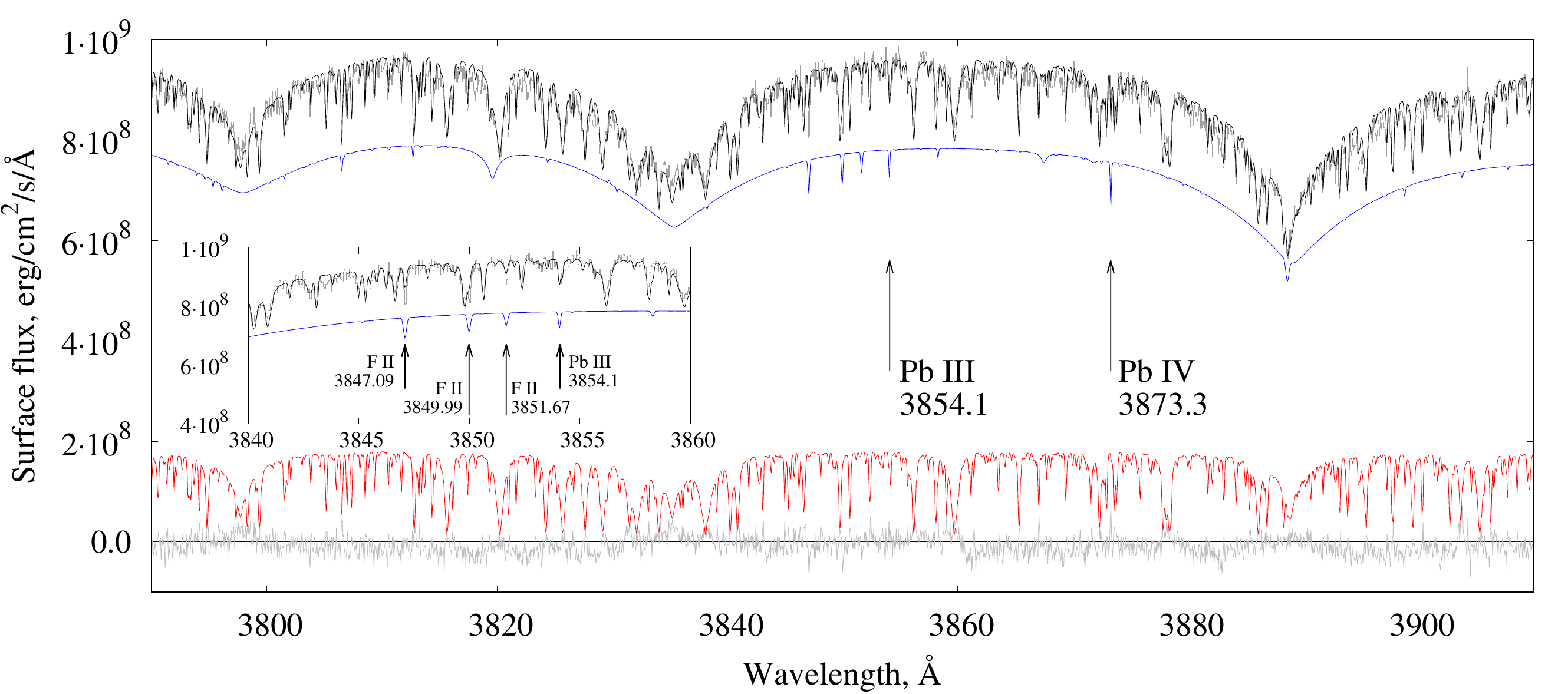}
\end{minipage}\\
\begin{minipage}{\textwidth}
  \centering
  \includegraphics[width=0.95\linewidth]{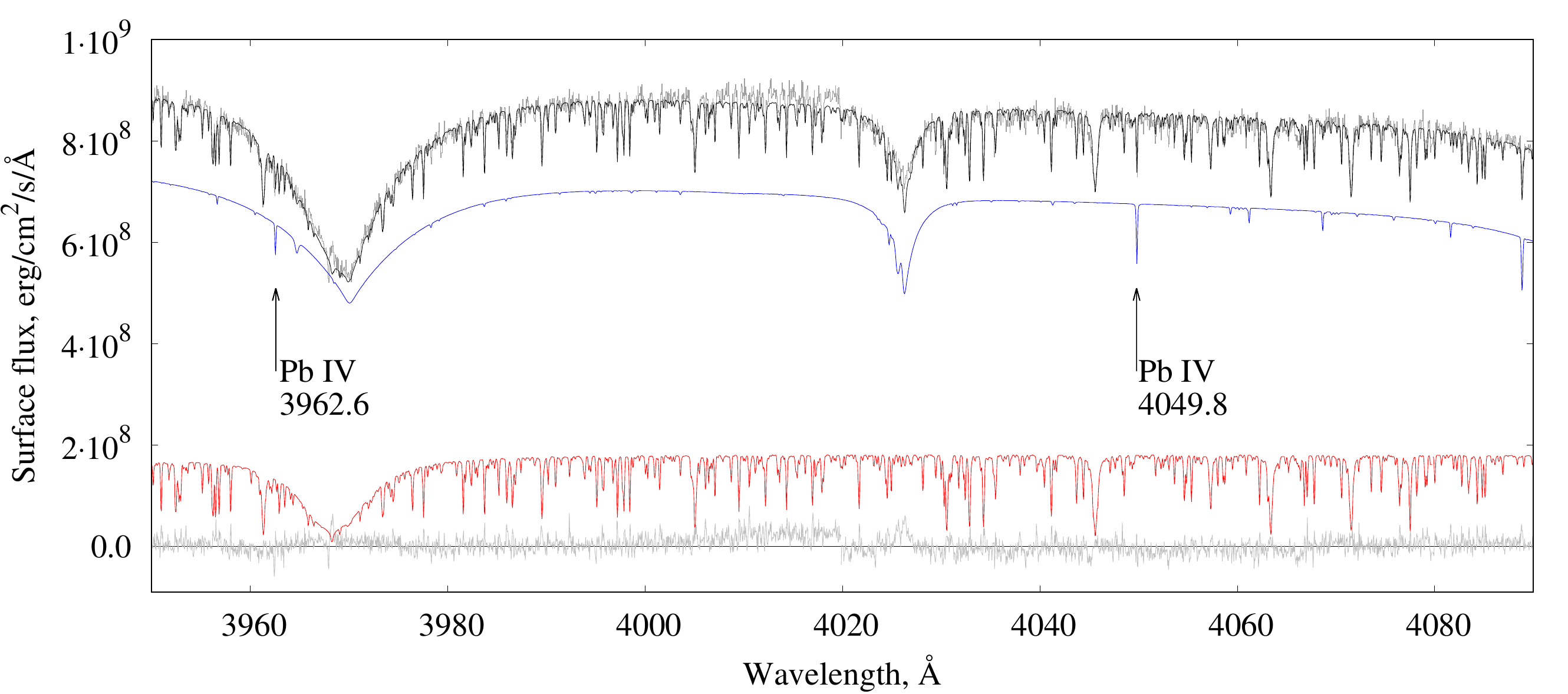}
\end{minipage}\\
\begin{minipage}{\textwidth}
  \centering
  \includegraphics[width=0.95\linewidth]{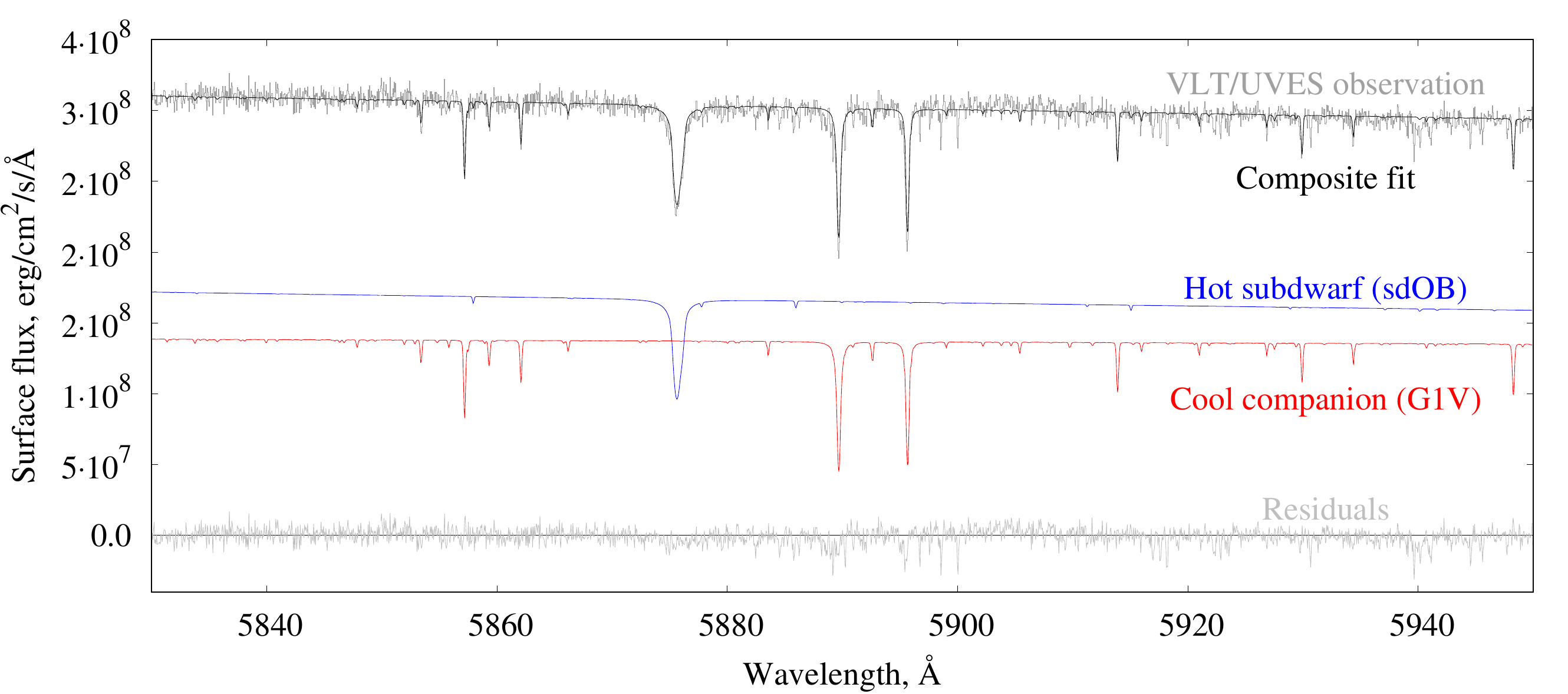}
\end{minipage}
  \caption{Selected regions of the best-fit {\sc Tlusty}/\xtgrid\ model of \mct. Strong lines of F and Pb are marked and labelled in the figures. The decomposition over the entire VLT/UVES spectral range is available online at: \mbox{\url{https://astroserver.org/TC43IV/}}
    \label{fig:xtgrid1}}
\end{figure*}

\begin{figure*}[]
  \centering
  \includegraphics[width=\linewidth]{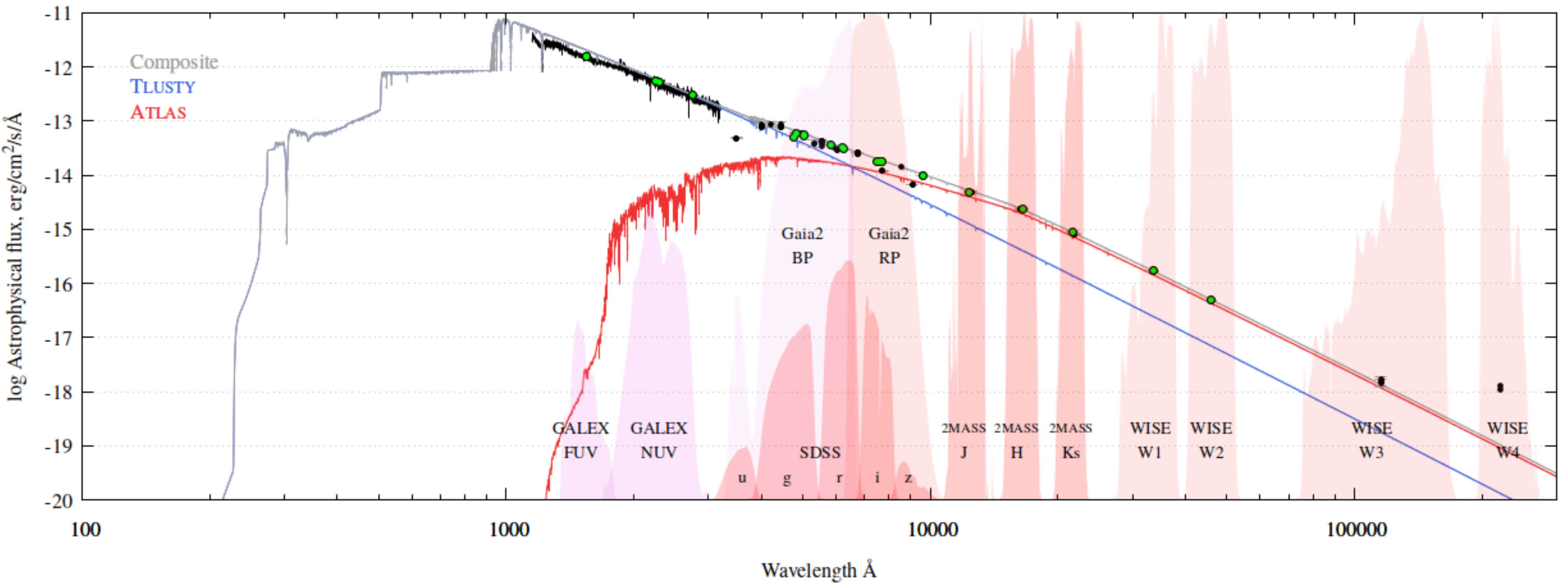}
  \caption{{\sc Tlusty}/\xtgrid\ spectral energy distribution. 
  The green data points are taken from the VizieR photometric data service, and used in the spectral decomposition to constrain the dilution factor.  
  Over the GALEX FUV and NUV ranges we show the observed IUE spectra (black). 
  The composite model (grey) is the sum of the {\sc Tlusty} sdOB (blue) and the {\sc Atlas} G1V (red) models. 
  The shaded regions mark various filter pass-bands.
    \label{fig:xtgrid2}}
\end{figure*}

Fig.\ref{fig:xtgrid1} shows selected regions of the best fit {\sc Tlusty/XTgrid} model to the UVES observations of \mct. 
Fig.\ref{fig:xtgrid2} shows the SED fit, which was used to improve the dilution factor in the spectral decomposition. 
The broad coverage of the SED shows a remarkable consistency from the UV to the far infrared regions. 
The surface parameters of \mct\ are found in Table \ref{tab:1}. 

The orbital phase resolved observations allow to separate the lines of the components based on their relative Doppler-shifts. 
This procedure also allows to identify spectral features that move with one of the components, even if the spectral models cannot reproduce them, or they are below the noise level of a single observation. 

The surface abundances of hot subdwarfs can be generally described by subsolar helium and light metal abundances, while iron is usually solar. 
In stars where trans-iron elements are detected, their abundances usually exceed the solar values by orders of magnitudes, most likely due to diffusion processes. 
Such an abundance pattern describes our findings well in \mct, with three noteworthy exceptions: 
\begin{enumerate}
\item While the light metal abundances are consistent with a tenth solar abundance, the iron abundance is low. 
This is consistent with the low metallicity of the companion and suggests that the binary formed in a metal poor environment of an old population.
\item We found strong lines of F in the spectrum of the subdwarf, corresponding to about 1000 times the solar F abundance. 
To our knowledge this is the first detection of F in a hot subdwarf star. 
\item We found strong Pb\,{\sc iii}-{\sc iv} lines that correspond to about 100\,000 times the solar lead abundance, but did not find similar, remarkably strong traces of other trans-iron elements.  
\end{enumerate}{}

The sdOB shows clear signs of a stratified atmosphere due to atomic diffusion: the wings of He lines are broad due to pressure broadening in deep layers and the central core absorption is shallow. 
This marks that He is mostly concentrated at the bottom of the atmosphere. 
Likewise, Pb lines that form near the photosphere are strong, while all expected strong lines that form higher in the atmosphere are absent. 
The strongest F lines are narrow in the observation while a homogeneous model produce too broad lines, indicating that most of F is likely concentrated above the photosphere and only the narrow line components are observed. 

\begin{table*}
\begin{center}
\caption{Spectroscopic surface and binary parameters from \xtgrid.
\label{tab:1}}
\begin{tabular}{lcrr}
\hline\hline
Primary parameters:    &&&   \\
$T_{\rm eff}$       & 37140 $\pm$ 850 K   & &  \\
$\log{g}$           & 5.588 $\pm$ 0.098 cm\,s$^{-2}$ & &  \\ 
Projected rotation  & < 3 km\,s$^{-1}$               & &  \\  
Chemical composition:      &&&      \\
Element (X)      &  Abundance        &Mass fraction&Solar fraction \\
                 &  $\log{n{\rm X}/n{\rm H}}$  & {\bf ${m_X/m_{\rm Total}}$}   & $\log{\epsilon/\epsilon_\odot}$ \\
H (reference)    &      0         &     9.18e-01 & 0.03  \\
He               &-1.69$\pm$0.09  &     7.41e-02 &-0.59  \\
C                &<-5.2           &     <8.64e-06 &<-2.51  \\
N                &-4.74$\pm$0.61  &     2.07e-04 &-0.59  \\
O                &<-4.8           &     <1.05e-04 &<-1.80  \\
F                &-4.73$\pm$0.20  &     3.20e-04 & 2.73  \\
Ne               &<-4.1           &     <1.32e-04 &<-1.05  \\
Na               &-4.1$\pm$0.5    &     8.27e-05 & 0.38  \\
Mg               &-5.5$\pm$0.6    &     7.50e-05 &-1.04  \\
Al               &<-6.2           &     <1.27e-05 &<-0.71  \\
Si               &<-5.8           &     <5.62e-05 &<-1.14  \\
P                &<-6.2           &     <1.83e-05 & <0.43  \\
S                &<-5.3           &     <1.47e-04 &<-0.39  \\
Ar               &<-5.0           &     <3.53e-04 & <0.61  \\
K                &<-6.0           &     <1.41e-05 & <0.51  \\
Ca               &<-5.4           &     <1.32e-04 & <0.24  \\
Ti               &<-5.4           &     <1.49e-04 & <1.61  \\
V                &<-5.3           &     <2.29e-04 & <2.79  \\
Cr               &<-5.1           &     <3.33e-04 & <1.23  \\
Mn               &<-5.2           &     <3.19e-04 & <1.40  \\
Fe               &-5.4$\pm$0.5    &     1.18e-04 &-1.11  \\
Co               &<-5.5           &     <1.58e-04 & <1.51  \\
Ni               &<-4.8           &     <7.27e-04 & <0.94  \\
Cu               &<-5.8           &     <7.44e-05 & <1.95  \\
Zn               &<-4.8           &     <8.84e-04 & <2.62  \\
Ga               &<-5.6           &     <1.36e-04 & <3.32  \\
Ge               &<-4.7           &     <1.13e-03 & <3.61  \\
Sr               &<-6.6           &     <1.98e-05 & <2.55  \\
Y                &<-7.2           &     <5.38e-06 & <2.64  \\
Zr               &<-6.5           &     <2.47e-05 & <2.92  \\
Pb               &-5.1$^{+0.2}_{-0.8}$  &     1.33e-03 & 5.13  \\
\hline
Companion parameters: &&& \\
Dilution factor$^{a}$  at 8900 \AA\ & 0.632 $\pm$ 0.050   \\
$\lambda_{F_{\rm MS}=F_{\rm sd}}$  & 6270 $\pm$ 35 \AA  \\
$T_{\rm eff}$                  & 5980 $\pm$ 200 K      \\
$\log{g}$           & 4.70 $\pm$ 0.3 cm\,s$^{-2}$       \\
Metallicity [M/H]$^{b}$                 & -1.02 $\pm$ 0.15      \\
Alpha enhancement [$\alpha$/M]                 & 0.20 $\pm$ 0.05      \\
Projected rotation & 4.1$^{+1.2}_{-0.6}$ km\,s$^{-1}$  \\
\hline
\end{tabular}\\
\parbox[t]{0.67\textwidth}{
a: Dilution factor = (Secondary flux) / (Primary flux + Secondary flux) = $F_{\rm
MS}/F_{\rm Total}$\\
b: [M/H] = $\log{z}$, where $z$ is the abundance ratio to hydrogen by number of the total sum of all elements except helium and hydrogen}
\end{center}
\end{table*}

\subsubsection{Fluorine}
Fluorine is rare in optical stellar spectra. 
\cite{Bhowmick20} investigated the presence of F in hot EHe stars. 
They found F in six out of ten stars and a level of overabundance is about 1000 times solar, very similar to the value we found in \mct. 
They found from trends of F with C, O, and Ne abundances that He burning after a white dwarf merger can account for the observed overabundances of fluorine.
The helium white dwarf merger models of \cite{Zhang12b} also predict a F overabundance in EHe stars. 
Although hot subdwarfs are significantly less massive than EHe stars, products from similar nucelosynthesis processes can occur in their atmospheres. 
The main production site of F is the inter-shell region of asymptotic giant branch
(AGB) stars and dredge-up can bring it to the surface.
\cite{werner05} have discovered F in extremely hot post-AGB stars. 
Similar conditions occur temporarily in hot subdwarfs that experience shell sub-flashes right after core He-ignition. 
Then the $^{13}$C pocket produce F that may be mixed into the atmosphere \citep{lugaro04}.

Our line list holds 523 lines of F\,{\sc ii} with $\log({gf})>-2$ in the observed UVES spectral range. 
The strongest optical lines of F\,{\sc ii} expected at $\lambda$4103.51\,\AA\ and $\lambda4246.72$\,\AA, are not observed.
The observed three lines originate from the lowest energy level 2s$^2$2p$^3$(4S$^\circ$)3s (176 493.93 cm\,$^{-1}$) among the optical transitions. 
The same lines have been observed by \cite{Bhowmick20}.

\subsubsection{Lead}
Lead is produced as the end product of the decay of heavier $r$-process elements and also produced by the $s$-process.
Table \ref{tab:lead} lists the three most widely used \pbfour\ lines in bold font, that have been found in heavy-metal sdOB stars \citep{Naslim13}. 
To extend this list and also include \pbthree\ in the analysis we have taken atomic energy levels from the NIST\footnote{\hyperref[https://physics.nist.gov/PhysRefData/ASD/levels_form.html]{https://physics.nist.gov/PhysRefData/ASD/levels\_form.html}} database and updated the line list with oscillator strengths of \pbthree\ and \pbfour\ from \citet{AM09} and \citet{AM11}, respectively. 
There is a notable systematic offset between the oscillator strengths by \citet{Naslim13} and \citet{AM11}, while the relative oscillator strengths of the lines are consistent. 
This offset resulted in higher Pb abundances in our analysis. 
\citet{Naslim13} found that the Pb\,{\sc iv}
$\lambda3962.48$ and Pb\,{\sc iv} $\lambda4049.80$ lines are present in most stars while the Pb\,{\sc iv} $\lambda4496.15$ line is often missing. 
We found the same behavior in \mct, which is expected in stratified atmospheres. 
If the vertical element distribution is inhomogeneous and the strongest lines form near the photosphere, it can occur that the lines are strong near the Balmer jump, as well as in the ultraviolet, but entirely missing in layers where the continuum forms outside the Pb rich layer. 
By identifying lines one can map the extent of the Pb enhanced layer. 
However, this requires a large set of relatively strong lines and multiple ionization stages. 
Such conditions are typically available in the atmospheres of cool stars and require high quality ultraviolet and infrared observations. 

\begin{table}[]
\begin{center}
\caption{Atomic data for \pbfour.\label{tab:lead}}
\begin{tabular}{cp{0.1\textwidth}p{0.1\textwidth}c}
\hline\hline
Wavelength    & \cite{Naslim13} & \cite{AM11} & Remark\\
     \AA\     & $\log{(gf)}$& $\log{(gf)}$ &\\
\hline
3873.31       & -        & -0.620 &  weak \\ 
{\bf 3962.48} & -0.047   & -0.287 &  strong \\ 
{\bf 4049.80} & -0.065   & -0.266 &  strong \\ 
4081.63       & -        & -1.398 & not observed \\ 
4174.29       & -        & -0.444 & not observed \\ 
{\bf 4496.15} & -0.237   & -0.437 & not observed \\ 
4605.43       & -        & -0.991 & weak \\ 
\hline
\end{tabular}
\end{center}
\end{table}

Although our homogeneous models are able to simultaneously reproduce the relative strengths of Pb\,{\sc iii} and Pb\,{\sc iv} lines, they are unable to assure that the derived Pb abundance reflects the true value. 
The uncertainty of atomic data and unknown stratification of Pb in the atmosphere limits the reliability of these results and requires further work.

Microturbulence has similar effects on relative line strengths like extreme stratification. 
\citet{Jeffery19} explored a range of turbulent velocities, but microturbulence could not reproduce the observed line strength inconsistencies.

The simultaneous presence of F and Pb, and the lack of other $s$-process elements in the atmosphere provides an opportunity to investigate diffusion in \mct. 
Either these elements are overproduced for some reason, or diffusion is more efficient on these two elements and they sink much slower to subphotospheric layers. 
The Eddington luminosity fraction $\log(L/L_{\rm Edd})=-2.49$ has a typical value for sdB stars, it is unlikely that \mct\ would have a stellar wind now. 
If the sdOB had a hotter and more extended stage in the past, the conditions were more favorable for a selective, opacity driven wind, and provide radiative support to highly ionized (heavy-)metals.

\subsection{Spectral analysis of the companion with GSSP}
A spectroscopic analysis of the main sequence (MS) companion was performed using the Grid Search in a Stellar Parameters (GSSP) code \citep{Tkachenko2015}. 
GSSP uses the method of atmosphere models and spectrum synthesis, comparing the observations with each theoretical spectrum in the grid. 
For the calculation of synthetic spectra, the SynthV LTE-based radiative transfer code \citep{Tsymbal1996} and an interpolated grid of LTE Kurucz atmosphere models \citep{Kurucz1979} were used in this work.

GSSP optimizes 7 stellar parameters at a time:
Effective temperature ($T_{\rm eff}$), surface gravity ($\log{g}$), metallicity ([M/H]), micro-turbulent velocity ($v_{\rm micro}$), macro-turbulent velocity ($v_{\rm macro}$), projected rotational velocity ($v_{\rm r}\sin{i}$) and the dilution factor ($F_{\rm MS}/F_{\rm Total}$) of the MS star.
A grid of theoretical spectra was built from all possible combinations of these parameters, and GSSP compared them with respect to the observed normalized spectrum, using a $\chi^2$ merit function to match the synthetic spectra to the observations.

A master spectrum was obtained by shifting the UVES spectra of \mct\ to the rest velocity of the MS star. 
We used the wavelength interval of 5904-6518\,\AA\, and normalized the observations using a polynomial function fitted to hand picked continuum points. 
This red-wards range is a compromise between a high S/N and high contribution of the cool companion. 
Spectral regions (like 6270-6330 \AA) containing telluric lines were avoided. 
The dilution factor was treated as a wavelength-independent factor for the short range in consideration here. 
\citet{Vos2018a} compared this approach with other spectral analysis methods and found that for short wavelength ranges, the fixed dilution factor does not reduce the accuracy.

The interpolated LTE Kurucz models extend the available metallicity range to [Fe/H] = -2.5 dex. 
The values of micro and macro-turbulent velocities were fixed in GSSP as the signal to noise of the master spectrum is not high enough to accurately determine these parameters. 
Their values, $v_{\rm micro}=1.0\pm0.2$\ km\,s$^{-1}$ 
and $v_{\rm macro}=2.9\pm1.7$ km\,s$^{-1}$ were obtained using the calibrated relations of \citet{Bruntt2010} and \citet{Doyle2014} respectively. 

The initial setup of the parameters search covers wide ranges with large step size in order to assure that the global minimum is found and prevents an excessive computational cost.
From the analysis of this first coarse grid a smaller step-size setup is arranged.
The final parameter values and errors are obtained from this last performance by fitting a polynomial function to the reduced $\chi^2$ coefficients. 
The minimum defines the final parameter value and the cutting values at 1$\sigma$, provide the error intervals. 
The set of best-fit parameters summarized in Table\,\ref{AP:GSSP} points to a G type star.

\begin{table}[]
\caption{Atmospheric parameters determined with the GSSP code in the wavelength range 5904-6517\,\AA, excluding the 6270-6330\,\AA\, range. 
The errors are 1$\sigma$ errors based on the reduced $\chi^2$.}
\label{AP:GSSP}
\centering
\renewcommand{\arraystretch}{1.4}
\begin{tabular}{lrl}
\hline\hline
\noalign{\smallskip}
Parameter    &   GSSP    &   Error   \\
\hline
\noalign{\smallskip}
$T_{\rm eff}$ (K)                  &5792    &   $^{+305}_{-285}$   \\ 
$\log{g}$ (dex)                    &4.60    &   $^{+0.55}_{-0.55}$ \\ 
$v_{\rm r}\,\sin{i}$ (km\,s$^{-1}$)&5.0     &   $^{+1.4}_{-2.0}$   \\ 
\big[M/H\big] (dex)                &$-$1.09 &   $^{+0.20}_{-0.24}$ \\
$F_{\rm MS}/F_{\rm Total}$       &0.53    &   $^{+0.11}_{-0.09}$ \\ 
\hline
\end{tabular}
\end{table}

To obtain chemical abundances, some intervals along the master spectrum with characteristic lines of specific elements from bibliography were used.
For abundances obtained from single lines, a previous study of the dilution factor is developed based on a larger wavelength range containing the studied line. 
Then, this value is hold fixed in the line study.
Abundances were obtained for the following elements: Na, Mg, Si, K, Ca, Sc, Ti, V, Cr, Mn, Fe, Co, Ni, Cu, Zn, Sr, Y, Zr, Ba and Eu. 
No remarkable deviations are observed with respect to the global metallicity and the expected trends by Galactic chemical evolution. 

CNO abundances in the MS star are key to understand binary interactions. 
The carbon abundance was derived using the single {\sc C\,i} line at 8335.14\,\AA. 
The [C/Fe] abundance ratio is consistent with the Galactic C abundance trends observed in abundances studies \citep[e.g.][]{Mattsson2010A&A...515A..68M,Amarsi2019A&A...630A.104A}. 
We could not obtain a reliable nitrogen abundance from optical lines as the {\sc N\,i} 8683.40\,\AA\, line is weak at low metallicity \citep{Takeda2005}.
The oxygen abundance determination was based on the strong {\sc O\,i} triplet lines at 7771.94, 7774.17 and 7775.39\,\AA\, and {\sc O\,i} lines at 8446\,\AA. 
The [O/Fe] is over the solar ratio, but consistent with the $\alpha$-element enhancement predicted by Galactic chemical evolution and Galactic abundance trends observed in abundance studies \citep[e.g.][]{Bensby2014A&A...562A..71B,Amarsi2019A&A...630A.104A}.
The carbon and oxygen abundances are listed in Table \ref{CNO:GSSP}. 

\begin{table}[]
\caption{C and O abundances of the G1V type companion in \mct.}
\label{CNO:GSSP}
\centering
\renewcommand{\arraystretch}{1.4}
\begin{tabular}{cclc}
\hline\hline
\noalign{\smallskip}
Element & $\log{n{\rm X}/n{\rm H}}$ & Error (1$\sigma$) & {[X/Fe]} \\
\hline
\noalign{\smallskip}
C           &  -4.53    &  $^{+0.01}_{-0.01}$  & +0.05     \\  
O           &  -3.61    &  $^{+0.01}_{-0.12}$  & +0.66     \\ 
\hline
\end{tabular}
\end{table}

Typical lines to derive lead abundances in MS stars are the two sensitive Pb\,{\sc i} lines at 3683.46 and 4057.81\,\AA. 
The 3683\,\AA\, line is out from the range of the UVES BLUE arm spectrum, but the 4057.81\,\AA\, line can be used, however, it is often weak and blended. 
It was not visually detected in our spectrum, but we could derive an upper limit from it \citep{Roederer2010}.
The upper limit on the Pb abundance is $\log{n{\rm X}/n{\rm H}}= -10.39$. 
Hence, we did not find any sign of Pb enhancement in the MS companion. 
In addition, our visual inspection of the entire master spectrum, looking for over-enhancement of the theoretical Pb lines from the NIST database had a negative result.

We did not find any suitable optical spectral lines for neutral or ionized fluorine in the main sequence star to obtain the fluorine abundance. 
Fluorine abundance determinations use vibrational lines in the near-IR (2.3 $\mu$m) from the HF molecule in relatively cool stars, which cannot be done with our spectra.

\section{Spectral energy distribution}
An independent way to determine effective temperatures of both components is by using the photometric spectral energy distribution (SED) based on literature photometry. For \mct\ we used Stromgren photometry obtained at the Cerro Tololo observatory \citep{Graham1973}, Gaia DR2 photometry \citep{GaiaDR2, Riello2018, Evans2018}, APASS DR9 \citep{Henden2015}, 2MASS \citep{Skrutskie2006} and WISE W1 and W2 \citep{Cutri2012}. 
The photometry is summarized in Table\,\ref{tb:photometry}. By using the Gaia parallax \citep{Lindegren2018, Luri2018}, also the radii of both components and their luminosity can be derived. 
We used the distance of $483^{+16}_{-15}$ pc by \cite{Bailer18}, which is only slightly different than the distance obtained by simply inverting the Gaia parallax ($489\pm15$ pc). 
The reddening in the direction of \mct\ as determined from the dust maps of \citet{Lallement2019} is E(B-V) = 0.01 $\pm$ 0.01, which is very low and will have little influence on the fit. 
The surface gravity cannot be constrained by an SED fit, and is kept fixed at the values determined from spectroscopy. 
The error on the surface gravity is propagated in the SED fit.

\begin{table}[]
\caption{Photometry of \mct\ collected from Gaia, APASS, 2MASS, WISE and \citet{Graham1973}.}
\label{tb:photometry}
\centering
\begin{tabular}{lrrr}
\hline\hline
\noalign{\smallskip}
Band    &   Magnitude    &   Error   \\
        &  mag      &   mag \\\hline
\noalign{\smallskip}
Gaia2 $G $      &   12.2376  &   0.0003    \\ 
Gaia2 $BP$      &   12.1800  &   0.0034    \\ 
Gaia2 $RP$      &   12.1845  &   0.0008    \\ 
APASS $B $      &   12.290   &   0.017     \\ 
APASS $V $      &   12.286   &   0.033     \\ 
APASS $G $      &   12.229   &   0.018     \\ 
APASS $R $      &   12.419   &   0.005     \\ 
APASS $I $      &   12.533   &   0.010     \\ 
2MASS $J $      &   11.997   &   0.024     \\
2MASS $H $      &   11.739   &   0.026     \\
2MASS $KS$      &   11.767   &   0.030     \\
WISE $W1$       &   11.691   &   0.024     \\
WISE $W2$       &   11.711   &   0.022     \\
STROMGREN $U$   &   12.328   &   0.033     \\ 
STROMGREN $B$   &   12.368   &   0.012     \\ 
STROMGREN $V$   &   12.433   &   0.020     \\ 
STROMGREN $Y$   &   12.340   &   0.010     \\ 
\hline
\end{tabular}
\end{table}

To fit the SED, models from the (T\"ubingen non-LTE Model-Atmosphere package \citep{Werner2003} are used for the sdOB component, and Kurucz atmosphere model \citep{Kurucz1979} for the companion. The fit uses a Markov chain Monte-Carlo approach to find the global minimum and determine the error on the parameters. A more detailed description of the fitting process is given in \citet{Vos2013, Vos2017} and \citet{Vos2018a}.

\begin{figure}[]
    \includegraphics{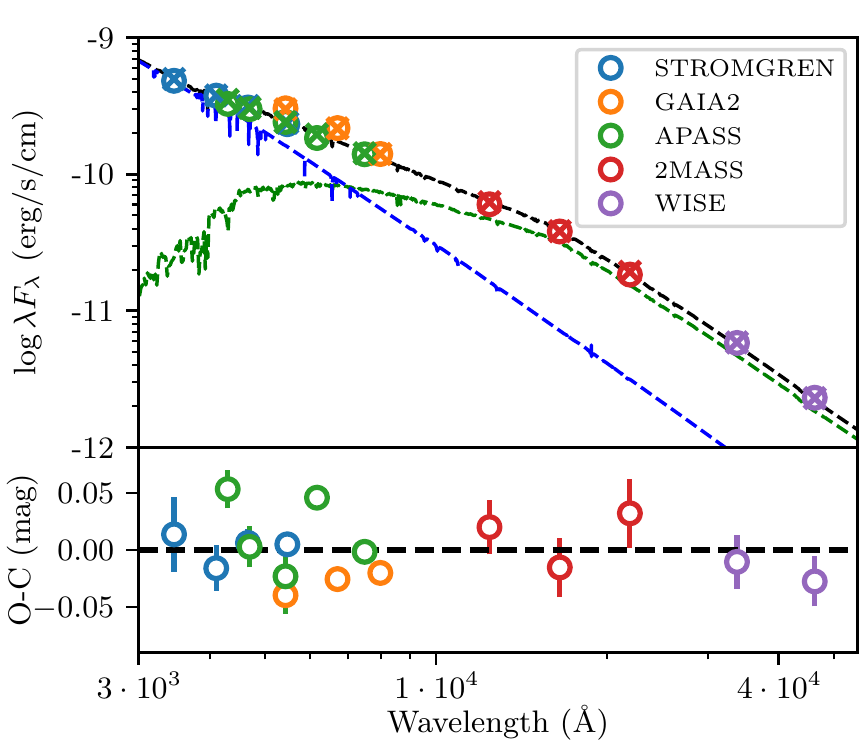}
    \caption{The SED of \mct\ together with the best fit model. The observed fluxes are shown as open circles while the calculated fluxes are shown as crosses. The best fit binary model is shown as a black dashed line with the cool and hot component shown respectively by green and blue lines.}
    \label{fig:sed_fit}
\end{figure}

The SED fit results in T$_{\rm eff}$ = 37\,000 $\pm$ 1800 K and R = 0.15 $\pm$ 0.01 R$_{\odot}$ for the sdOB component and T$_{\rm eff}$ = 5800 $\pm$ 240 K and R = 0.93 $\pm$ 0.03 R$_{\odot}$ for the cool companion. The total luminosity is 28.5 $\pm$ 2.2 L$_{\odot}$ and 0.82 $\pm$ 0.05 L$_{\odot}$ for the sdOB and companion respectively. These results are in very good agreement with the parameters determined from spectroscopy. In Fig.\,\ref{fig:sed_fit}, the photometry of \mct\ is shown together with the best fitting model and it's decomposition in the sdOB and MS component. The reduced $\chi^2$ value of the best fitting model is 4.1. This indicates that the errors on the photometry are likely a bit underestimated, which is a known problem in many photometric surveys. However, the $O-C$ values of the SED fit are random, and the good match between parameters determined from spectroscopy and those from photometry indicate that our spectroscopic parameters for \mct\ are reliable.

\section{Mass determination}
Masses of the hot subdwarf and the cool companion can be determined using the surface gravity determined from the spectroscopic analysis, and the radius from the photometric SED. 
However, the surface gravity is difficult to determine in composite spectra (e.g. \citealt{Nemeth16,Vos2018a}) because the blended lines cause a strong degeneracy between the components.
An alternative to using the surface gravity and radius is to use stellar evolution models. 
The different effects at play in the atmospheres of hot subdwarfs that all affect the observed spectroscopic parameters make this method difficult for sdOB stars, but the cool companions are regular main sequence stars. 
They can have accreted some mass of the sdOB, but this is not more than a few percent of a solar mass, and enough time has passed since the mass loss phase to allow the cool companion to settle again. 

Here we will follow a similar approach as was used in \citet{Maxted2015} to determine the mass of the cool companion in \mct. 
A Markov chain Monte-Carlo code is used to fit an evolution model to the effective temperature, radius, luminosity and metallicity of the cool companion. The stellar evolution models used here are the MESA Isochrones and Stellar Tracks (MIST) models version 1.2 \citep{Choi2016, Dotter2016} which are calculated using MESA version r7503 \citep{Paxton2011, Paxton2013, Paxton2015}. This method is the same as used and tested on the DEBcat \citep{Southworth2015} eclipsing binaries catalog in \citet{Vos2018b}.

\begin{figure}[]
\includegraphics{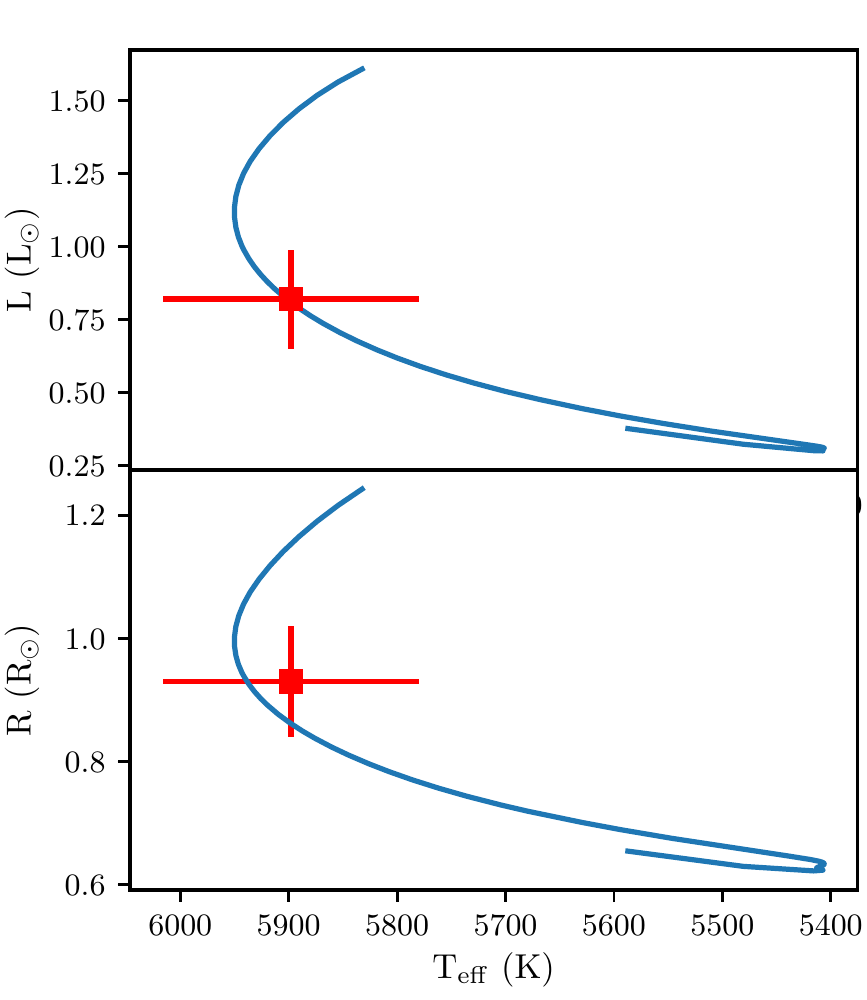}
\caption{The radius, luminosity and effective temperature of the cool companion in \mct\ plotted in red, with the best fitting MIST stellar track in blue. Only the main sequence part of the track is shown. This track fits the observed metallicity of the cool companion and has a mass of 0.72 M\sun.}
\label{fig:mass_determination}
\end{figure}

Using this method we find a mass of 0.72 $\pm$ 0.05 M\sun\ for the cool companion. Using the mass ratio derived from the spectroscopic orbit \citep{Vos2019a}, the mass of the sdOB star is 0.47 $\pm$ 0.06 M\sun. This mass is very close to the canonical sdB mass of roughly 0.5 M\sun. The best fitting track to the radius, luminosity and effective temperature of the cool companion is shown in Fig.\,\ref{fig:mass_determination}. 
With the mass of the sdOB being almost exactly the same as the canonical mass, this sdOB is likely formed from a low mass progenitor that ignited He under degenerate circumstances. While sdBs that form from higher mass progenitors and ignite He in a non-degenerate core can have similar core masses, the majority of such systems have masses that are lower that those formed from low mass progenitors \citep{Han2002}.

\section{Galactic orbit}
\begin{figure}[]
    \includegraphics[width=\linewidth]{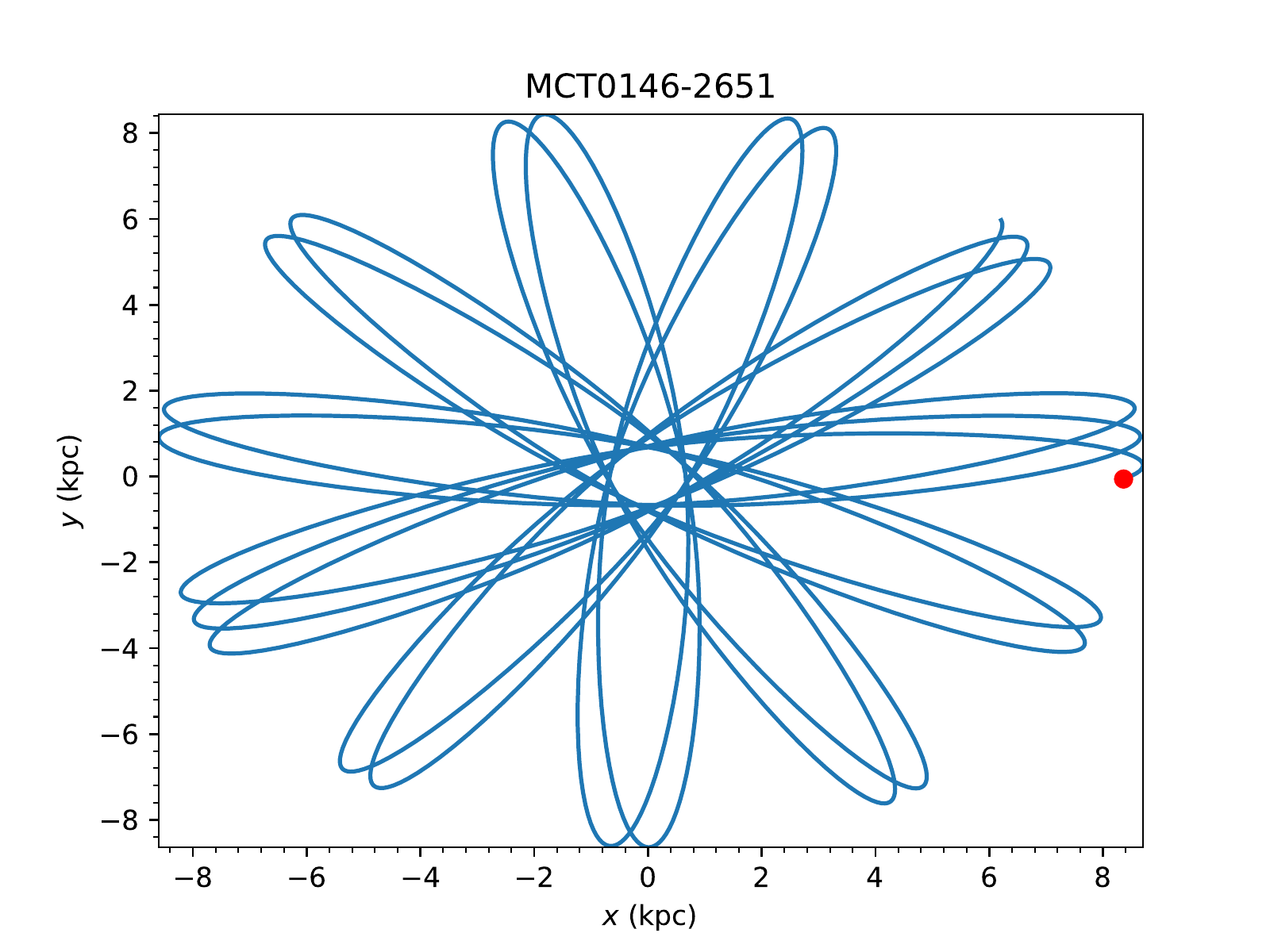}
    \caption{The Galactic orbit of \mct\ projected onto the Galactic plane. The trajectory over a time of 2 Gyr and the current position (red dot) is shown.}
    \label{fig:galactic_orbit}
\end{figure}

We used the {\sc galpy} software package \citep{Bovy2015} to calculate the kinematic properties and Galactic orbit of \mct\ in the Galactic potential. 
The important parameters here are the Galactic velocity components U, V, and W, as well as the orbital eccentricity $e$ and the angular momentum component perpendicular to the Galactic plane $J_z$. 
The uncertainties were estimated via a Monte Carlo process. 
The calculated orbit in the Galactic X-Y plane can be seen in Fig.\,\ref{fig:galactic_orbit}.

While the velocity components of \mct, U = 21.28 $\pm$ 0.09 km s$^{-1}$, V = 232.86 $\pm$ 0.07 km s$^{-1}$, W = -38.97 $\pm$ 0.5 km s$^{-1}$ are not enough on their own to distinguish it from a thin/thick disk star, the angular momentum and eccentricity of the Galactic orbit help to differentiate from the disk population. 
An average thin disk star has an eccentricity less than 0.3, and an angular momentum close to 1800 kpc km s$^{-1}$, thick disk stars have an eccentricity below 0.75, and an angular momentum above 500 kpc km s$^{-1}$ \citep{Luo20}. 
\mct\ has a very high eccentricity of 0.89 $\pm$ 0.02 and very low angular momentum of 237 $\pm$ 55 kpc km s$^{-1}$, placing it firmly in the Halo population. 
This is further supported by the low metallicity of the system of [Fe/H] = -1.02 $\pm$ 0.04 dex. 

The few known heavy-metal hot subdwarfs show a kinematic pattern. 
While most of the He-rich sdOB stars belong to the thin disk \citep{Martin17}, heavy metal sdOB stars belong to the Halo population \citep{Dorsch20}. 
We find \mct\ consistent with this general picture. 

\section{Comparison with other long period sdB binaries}
\begin{figure*}[]
    \includegraphics{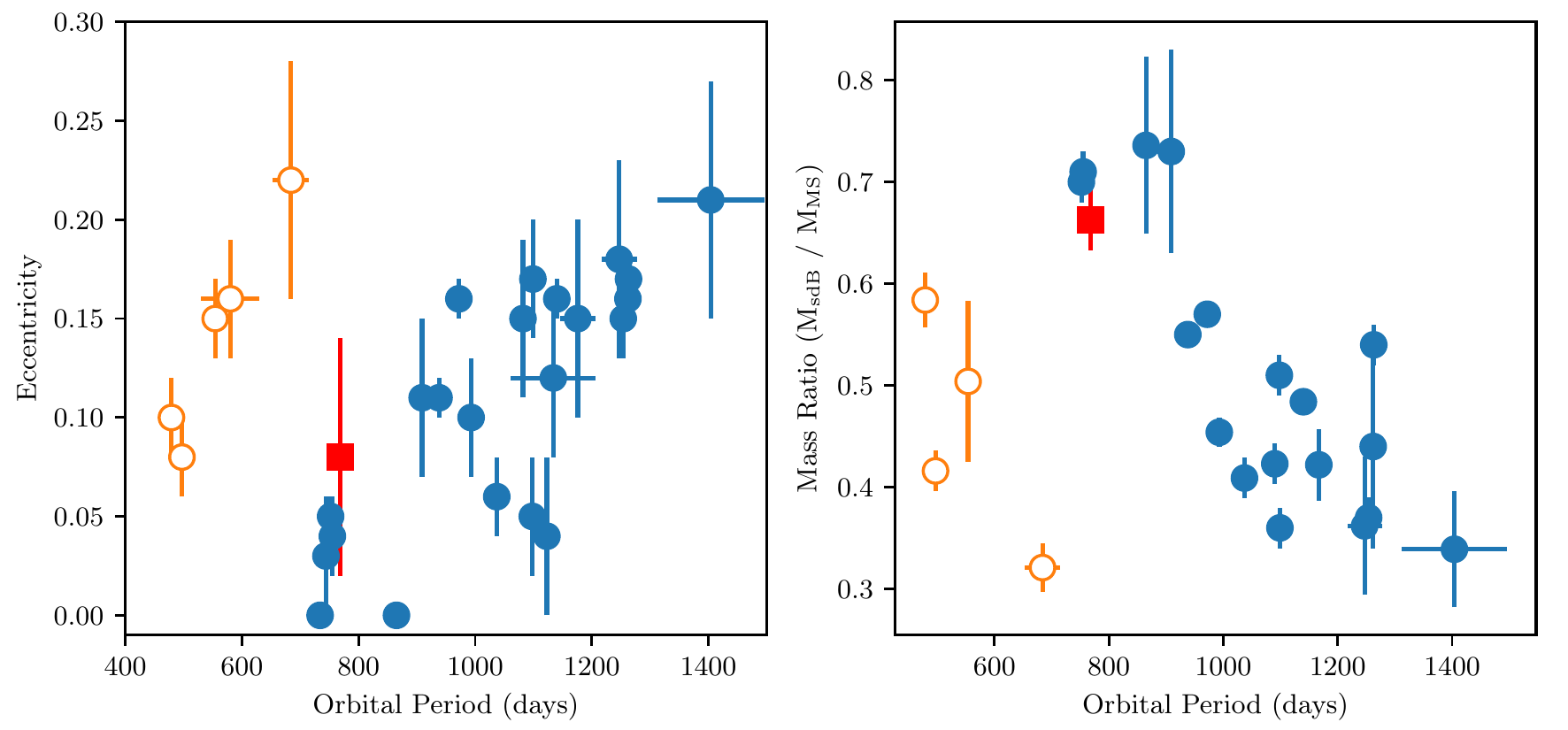}
    \caption{The orbital period versus the eccentricity and mass ratio for all known composite sdB binaries with solved orbits. The main group is shown in blue filled circles, the secondary group in orange open circles, and \mct\ is highlighted as a red square. The orbital parameters of \mct\ fit with those of the main group.}
    \label{fig:mct_composite_binaries_comparison}
\end{figure*}

Currently there are 25 long period composite sdB binaries known with solved orbital parameters, of which 23 also have known mass ratios. 
See e.g. \citet{Vos2019b} or \citet{Vos2020} for an overview of these systems. 
In Fig.\,\ref{fig:mct_composite_binaries_comparison} the orbital period, mass ratio and eccentricity of all long period composite sdB binaries is plotted with \mct\ highlighted in red. 
These systems can be subdivided in two groups, the main group containing most systems, is shown in blue. 
The second group, which contains only a few systems, is shown in orange. 
Both groups show the same correlations of higher eccentricity at longer orbital period and lower mass ratio at wider orbits. 
However, the second group is found at shorter orbital periods. 
\citet{Vos2020} has found that the main group of systems can be formed from main sequence binaries with primary masses between 0.8 and 1.8 M\sun, with orbital periods varying between 100 and 900 days. 
The P-q correlation is explained by Galactic evolution. 

When comparing the orbital parameters of \mct\ with the known composite sdB binaries, \mct\ falls in the main group of systems. 
There are no signs that this system is an outlier, neither based on the orbital characteristics, nor based on its mass. 
This would indicate that \mct\ is likely formed in the same way as the majority of long period hot subdwarf binaries have formed.

\section{Discussion and conclusions}
The results support that a combination of mixing due to either binary evolution (merger) or single star evolution (dredge-up) and atomic diffusion are responsible for the observed F and Pb overabundances. 
Either a stellar merger or flash induced mixing brings core material to the surface, and diffusion supports highly ionized elements for longer periods of time.  
While atomic diffusion erases all metallicity based chemical history of the subdwarf older than a few million years, thanks to the composite nature of \mct, the companion reveals the age and past interactions of the members, which put constraints on the viable formation channels. 
The stars we see with heavy-metals are likely not exceptional and follow normal stellar evolution tracks, but we catch them in a particular phase when temporarily they are enriched in $s$-process elements. 

While Pb has been observed in a dozen of sdOB stars, \mct\ is a first such star that shows a measurable F abundance. 
Fluorine is not often seen in optical spectra of stars, hence its Galactic distribution is not known well. 
It is synthesized in three major sites: supernovae, inter-shell zones of AGB stars and Wolf-Rayet stars.
We associate the observed F abundance with He burning and atmospheric mixing. 
A strong bimodality of the F/Na ratio has been observed in globular clusters.
This is consistent with stellar generations, where first generation stars are F-rich and Na-poor and second generation stars are F-poor and Na-rich \citep{Laverny13}. 
This observation also suggest that \mct\ is an old system. 

\citet{Vos2018a} have shown that distant companions accrete only a few per cent of a solar mass during the Roche-lobe overflow stage, not enough to significantly change their evolution time scales. 
Therefore the age of the companion in \mct\ represents the age of the system at the accuracy of stellar evolution models. 
Assuming that \mct\ has a young subdwarf, this age also marks how much time the sdOB progenitor had to reach its current phase. 
The subdwarf phase is a relatively short period compared to stellar evolution, which puts constraints on the initial mass of the subdwarf progenitor because the core mass depends on the initial mass, and the time spent on the RGB. 
Or, in case of a merger, a constraint can be given on the total mass of the components. 
The problem is that both binary star and triple star evolution can lead to the currently observed sdOB+MS binary.

It is not possible to uncover the history of the subdwarf in \mct\ from its current state and available data. 
Both single star evolution and binary merger models are able to explain the current  observations.

If a single star evolution is considered: 
It is possible that a low-mass (1 - 2 M\sun) star evolved through the RGB and experienced mass-loss through Roche-lobe overflow. 
The mass loss interrupted further evolution on the RGB. 
The stripped He core of the subdwarf progenitor started its evolution on the white dwarf cooling sequence and experienced a delayed core He flash.
Following the ignition of the He core, these stars experience a sequence of shell He-flashes, when the envelope material is mixed with new $s$-process elements from hot bottom burning. 
Finally the star settles in its new equilibrium and slowly reaches its position corresponding to its core and envelope mass combination on the zero-age EHB. 
In the calm atmosphere of EHB stars atomic diffusion is the major process that governs the observable abundance pattern. 
The spectral decomposition suggests that the companion is G1V type, a solar-like star, and the binary orbit shows its mass is 0.73 M\sun corresponding to an early K type. 
This discrepancy is due to the low metallicity of the companion that increases the $T_{\rm eff}$ corresponding to the same mass. 
Therefore, a Population II Halo MS star with 0.8 M\sun\ appears to be an early G type star, similar to a 1 M\sun\ Population I star \citep{GaiaDR2}. 

The binary star scenario starts with a hierarchical triple system with a more massive inner binary (0.8 - 2 M\sun). 
Depending on the initial masses, the inner binary may pass through one or two common-envelope stages. At the end of the common envelope evolution the stars merge, ignite He core burning and form a hot subdwarf. 
If the stars do no merge at the end of the common envelope stage a close
white dwarf binary remains that loses orbital angular momentum and cannot avoid the final merger. 
The distant companion is not involved in the formation of the subdwarf. 
The binary merger channel also requires an efficient method to remove the angular momentum, as the subdwarf does not show a measurable projected rotation.  
The observed $[{\rm Fe}/{\rm H}]=-1.0$ dex and $[\alpha/{\rm Fe}]=0.2$ of the companion suggests that system is likely older than 11 Gyr, which allows plenty time to evolve through orbital decay and a merger.  

The example of \mct\ shows that chemically homogeneous models are inappropriate for precision analyses of heavy-metal stars, yet important to provide first estimates. 
Although stratified atmosphere models, such as {\sc Tlusty}, are already available, the efficient fitting of such models to observations is a demanding task just like to obtain the necessary high quality observations. 

The mass of the sdOB is fully consistent with EHB stars, but its location in the $T_{\rm eff}-\log{g}$ plane is not. 
This means that the star is either young and evolves towards the EHB, or already passed the EHB and now is in a post-EHB He-shell burning stage. 
None of the post-EHB stars show any measurable heavy-metal abundances, therefore we conclude that \mct\ must have a young sdOB star that has not reached the EHB yet. 
An alternative scenario would be if stars at the end of the EHB stage would experience a dredge up that mixes the envelope. 

Main points: 
\begin{itemize}
    \item the sdOB in \mct\ has observed overabundances of Pb and F. 
    \item the companion does not show extra overabundances, and looks like a normal low metallicity MS star. The abundance pattern from the current data does not show past mass accretion from the sdOB progenitor.
    An [$\alpha$/M] = $0.25\pm0.05$ dex alpha element enhancement is observed, which is consistent with Population II stars.
    \item the mass of the sdOB is consistent with the canonical mass and is likely formed from a low-mass progenitor, or a merger.
    \item the Galactic orbit points that \mct\ is a Halo object, like all other heavy-metal hot subdwarfs discovered to date. 
    \item \mct\ is a typical composite binary with orbital parameters similar to those of other composite subdwarf binaries. 
    There is no reason to assume that its formation history was different from other composite hot subdwarfs.
    \item the existence of the companion rejects the 'companion explosion' formation channel, and shows that the heavy-metal overabundances are not primordial or caused by an external source, but are pure atmospheric effects. 
    \item \mct\ demonstrates that heavy-metal overabundances occur in metal poor stars, where efficient atomic diffusion can support the F and Pb abundance in thin photospheric layers. 
    \item the period of atmospheric overabundances are probably temporary, short lived phases compared to stellar evolution. 
    This explains why not all similar stars show heavy-metal overabundances, which are most likely in connection with mixing events following a He-shell subflash \citep{Marcelo08}. 
    Therefore, we expect the number of stars with overabundances to anticorrelate with the time since the last subflash event.
    \item the sdOB has a low iron abundance and an alpha element enhancement, which are in line with its assumed low metallicity, Population II origin.
\end{itemize}
We conclude that \mct\ is a unique system, which takes the investigations of heavy-metal subdwarfs to a new level and implies, that more such subdwarfs may be hidden in composite spectrum binaries.
Ultraviolet spectroscopic observations and stratified atmosphere models will be necessary to investigate the complete chemical profile of heavy-metal hot subdwarfs and reproduce their observed features.

\begin{acknowledgements}
PN acknowledges support from the Grant Agency of the Czech Republic (GA\v{C}R 18-20083S). 
This work was supported by a fellowship for postdoctoral researchers from the Alexander von Humboldt Foundation awarded to JV.
This research has used the services of \mbox{\url{www.Astroserver.org}} under reference TC43IV.
This research has made use of the VizieR catalogue access tool, CDS, Strasbourg, France.
Based on observations collected at the European Southern Observatory under ESO programmes 088.D-0364, 093.D-0629 and 0104.D-0135.
This work has made use of data from the European Space Agency (ESA) mission {\it Gaia} (\url{https://www.cosmos.esa.int/gaia}), processed by the {\it Gaia} Data Processing and Analysis Consortium (DPAC, \url{https://www.cosmos.esa.int/web/gaia/dpac/consortium}). 
Funding for the DPAC has been provided by national institutions, in particular the institutions participating in the {\it Gaia} Multilateral Agreement.
\end{acknowledgements}

\bibliographystyle{aa}
\bibliography{bibliography.bib}

%
%

\end{document}